\pdfoutput=1
\documentclass[sigconf]{acmart}

\acmConference[HATRA '24]{Human Aspects of Types and Reasoning Assistants}{20 October 2024}{Pasadena, CA, USA}

\acmYear{2024}
\acmMonth{10}
\acmDOI{} 
\startPage{1}

\setcopyright{none}

\usepackage{booktabs}   
\usepackage{subcaption} 
\usepackage{xargs}
\usepackage{todonotes}
\newcommandx{\done}[2][1=]{\todo[linecolor=blue,backgroundcolor=blue!25,bordercolor=blue,#1]{#2}}

\usepackage{listings}
\usepackage{xcolor}

\definecolor{codegreen}{rgb}{0,0.6,0}
\definecolor{codegray}{rgb}{0.5,0.5,0.5}
\definecolor{codepurple}{rgb}{0.58,0,0.82}
\definecolor{backcolour}{rgb}{0.95,0.95,0.92}

\lstdefinestyle{mystyle}{
    backgroundcolor=\color{backcolour},   
    commentstyle=\color{codegreen},
    keywordstyle=\color{magenta},
    numberstyle=\tiny\color{codegray},
    stringstyle=\color{codepurple},
    basicstyle=\ttfamily\footnotesize,
    breakatwhitespace=false,         
    breaklines=true,                 
    captionpos=b,                    
    keepspaces=true,                 
    numbers=left,                    
    numbersep=5pt,                  
    showspaces=false,                
    showstringspaces=false,
    showtabs=false,                  
    tabsize=2
}

\begin{document}

\title{Surveying the Rust Verification Landscape}         


\author{Alex Le Blanc}
\affiliation{
  \institution{University of Waterloo}            
  \country{Canada}                    
}
\email{a6leblan@uwaterloo.ca}          

\author{Patrick Lam}
\affiliation{
  \institution{University of Waterloo}            
  \country{Canada}                    
}
\email{patrick.lam@uwaterloo.ca}          

\begin{abstract}
  Rust aims to be a safe programming language applicable to systems
programming applications. In particular, its type system has strong guardrails to prevent a variety of issues, such as memory safety bugs and data races. However, these guardrails can be sidestepped via the \verb|unsafe| keyword. \verb|unsafe| allows certain otherwise-prohibited operations, but shifts the onus of preventing undefined behaviour from the Rust language's compile-time checks to the developer. We believe that tools have a role to play in ensuring the absence of undefined behaviour in the presence of unsafe code. Moreover, safety aside, programs would also benefit from being verified for functional correctness, ensuring that they meet their specifications.

In this research proposal, we explore what
it means to do Rust verification. Specifically, we explore which
properties are worth verifying for Rust; what techniques exist to
verify them; and which code is worth verifying. In doing so, we motivate
an effort to verify safety properties of the Rust standard library, presenting the relevant challenges along with ideas to address them.


\end{abstract}

\ccsdesc[500]{Software and its engineering~General programming languages}
\ccsdesc[300]{Social and professional topics~History of programming languages}


\maketitle

\newcommand{\ie}{\textit{i.e.,} }
\newcommand{\cf}{\textit{cf.} }
\newcommand{\eg}{\textit{e.g.,} }
\newcommand{\etal}{\textit{et al.}}

\newcommand{\gil}[1]{{\textcolor{blue}{\textbf{Gil: #1}}}}
\newcommand{\jeehoon}[1]{{\textcolor{green!60!black}{\textbf{Jeehoon: #1}}}}
\newcommand{\yoonseung}[1]{{\textcolor{pink!40!black}{\textbf{Yoonseung: #1}}}}
\newcommand{\youngju}[1]{{\textcolor{pink!60!black}{\textbf{YoungJu: #1}}}}
\newcommand{\juneyoung}[1]{{\textcolor{brown}{\textbf{Juneyoung: #1}}}}
\newcommand{\sanghoon}[1]{{\textcolor{brown!60!black}{\textbf{Sanghoon: #1}}}}
\newcommand{\dongyeon}[1]{{\textcolor{brown!40!black}{\textbf{Dongyeon: #1}}}}


\section{Introduction}
\label{sec:introduction}

Rust is a modern programming language which aims to be ergonomic and suitable for high-performance systems programming while making it harder to write bugs.
As the language designers put it, as quoted in the Rust Book~\cite{klabnik24:_rust_progr_languag}:
\begin{quote}
  The Rust programming language helps you write faster, more reliable software.
\end{quote}
Specifically, Rust has a number of features that are novel to widely-deployed programming languages, notably its ownership-based type system used for memory management and to guarantee race freedom. Multi-million line Rust codebases exist; even in 2020, the Servo and TiKV projects exceeded 2 million lines of code~\cite{anderson20:_rust_compil_model_calam}, in addition to the Rust compiler itself (another million lines). Unlike C/C++, Rust aims to limit undefined behaviour to code specifically marked as unsafe; and, unlike Java, Rust does not pretend that unsafe code does not exist.

Amazon is interested in verifying the Rust standard library\footnote{\url{https://model-checking.github.io/verify-rust-std/intro.html}} and aims to incite community participation. While the Rust authors' goal is to enable developers to write ``faster, more reliable'' software, we are specifically interested in contributing techniques to this effort that strengthen specific aspects of Rust's ``more reliable'' aspiration and understanding human factors related to carrying out this verification.

In this work, our goal is to sound out the surrounding landscape. We explore a programme for ``verifying Rust''---what might it mean to verify some nontrivial body of Rust code? Does that mean that Rust code should not crash? Or should it have certain specific functional correctness properties? Are we talking about all Rust code, or just code written in Safe Rust? Is the standard library a reasonable place for us to start? What are the advantages and disadvantages of Amazon's Kani tool and others in its space? What exists in other languages?

Specifically, in this research proposal, we consider three fundamental
questions: what properties should be verified; how should they be
verified; and what code should be verified? 

We start (Section~\ref{sec:properties}) by exploring which
properties are worth verifying. Rust generally aims to prevent memory
safety errors and race conditions, but as a systems programming
language, it also explicitly allows programmers to use Unsafe Rust,
bypassing safeguards. We also discuss
properties beyond generic safety properties, namely functional
correctness or domain-specific properties. Although our goal is to
verify Rust, we also survey (Section~\ref{sec:tools}) the situation of
program verification in C/C++/Java, as well as existing tools for
verifying Rust and the techniques they use.

We also investigate (Section~\ref{sec:which-code}) which code is worth
verifying. Rust applications rely on libraries, and almost always on
the Rust standard library. Applications, libraries, and the standard
library have different verification needs: one might expect that
verifying domain-specific properties is more relatively valuable for
application code than verifying memory safety of unsafe code.
The standard library, by contrast, is a heavy user of unsafe code,
and its safety is assumed by the vast amounts of code that depend on it,
so that verifying memory safety is especially important there.
We then describe (Section~\ref{sec:challenges}) our proposed approach
to verifying Rust and anticipated challenges. Finally, we discuss (Section~\ref{sec:human}) additional long-term objectives related to human factors, beyond just the verification of Rust code.



\section{Rust, Properties, and Unsafe Code}
\label{sec:properties}

We first explore the question of what properties we focus
on.  One could take an empirical approach and focus on techniques that
address the most common types of Rust bugs, or perhaps the most
important bugs; in a related vein, \citet{shirzad24:_rust} investigated
Rust bug fix patterns, and specifically those related to borrowing.

However, one of Rust's primary design goals is to be a ``safe'' programming
language, while also supporting low-level features needed for systems
programming. Thus, per
the Rustonomicon~\cite{rustonomicon} (part of Rust's official documentation), the subset of Rust known as Safe Rust is designed to
prevent undefined behaviour, providing various safety guarantees.
For instance, by design, Safe Rust programs ought to be memory safe and free of data races.
The Rustonomicon includes the sentence:
\begin{quote}
  Safe Rust is the true Rust programming language.
\end{quote}

Unsafe Rust differs from Safe Rust in that undefined behaviour can happen.
Because Safe Rust is free of undefined behaviour, even buggy
programs follow Rust semantics.  On the other hand, there are no
semantics for any Unsafe Rust code that executes undefined
behaviour. Thus, one of the primary goals behind the research
described in this proposal is to eliminate all undefined behaviour, by
verifying Unsafe Rust.  This then enables sound reasoning about the
behaviour of the entire program.  In particular, we plan to start
by verifying the Unsafe Rust in the standard library.

As a caveat, to our knowledge, the ``no undefined behaviour'' guarantee of Safe Rust has not been
formally proven.
An effort analogous to the CompCert
project~\cite{Leroy-Compcert-CACM} would be required to formally show the
absence of implementation bugs in the specification and compiler.
Limiting the scope of such an effort to Safe Rust is likely to be simpler
than aiming to verify an optimizing C compiler, as in CompCert's case.
However, there is empirical evidence that suggests that Safe Rust's guarantees hold in practice. \citet{xu21:_memor} report, from a study of 186 Rust CVEs,
that undefined behaviour in their Rust programs
always came from bugs in Unsafe Rust code, rather than
Safe Rust.

For the work that we are proposing, we believe that it is reasonable to assume that
(putting aside language specification and
implementation bugs) a Rust programmer must use Unsafe Rust to
get undefined behaviour---either directly, or by interacting with
Unsafe Rust. In contrast to Safe Rust, Unsafe Rust
explicitly includes constructs that can trigger undefined behaviour if
used inappropriately.

Contrast Rust's situation with C and C++, where any line of code may
inadvertently trigger undefined behaviour. Rust has a significant
advantage here: it is feasible to audit Rust code for the program
locations which may contain undefined behaviour, because such
locations must be associated with some usage of the \texttt{unsafe}
keyword.

As an example of undefined behaviour in Rust, the Rustonomicon sketches\footnote{\emph{Rustonomicon}, section 1.3, ``Working with
Unsafe'':
\url{https://doc.rust-lang.org/nomicon/working-with-unsafe.html}}
an implementation for a struct that we call \texttt{MyVec}.
This struct could include a
\texttt{make\_room} function (in Safe Rust) that modifies the
\texttt{MyVec}'s capacity, along with unsafe code that relies on the
capacity being accurate. A client with access to \texttt{make\_room}
can trigger undefined behaviour in the unsafe
code. \texttt{make\_room} does not have to be accessible by clients;
if it is inaccessible, then \texttt{MyVec}'s unsafety is encapsulated, all of its clients are sound, and
undefined behaviour is impossible. If it is accessible, then only
\texttt{MyVec} clients that use \texttt{make\_room} appropriately are
sound, and a verifier must check that all calls to that method
(1) meet necessary preconditions and (2) do not misuse returned values.

\paragraph{Encapsulating unsafety.}
Within a module that contains Unsafe Rust, undefined behaviour may be triggered by errors in the unsafe code, or by safe code that nevertheless has access to directly break invariants that the unsafe code depends on. As we have seen in the example above, it is possible for an Unsafe Rust-containing module to provide access to functions that break its invariants; clients of that module can---if they break the safety contract---indirectly break the invariants and cause unsafe behaviour, but they cannot break them directly.

Some libraries
\emph{encapsulate} unsafety, so that clients can never trigger undefined
behaviour. That is, for those libraries, the required safety conditions on the client are nil. As we have seen, though, other libraries (including the Rust standard library), do not.
\citet{ozdemir19:_unsaf_rust} names the phenomenon of poorly-encapsulated safety properties ``Abstraction Escape'' and proposes a static analysis to detect simple cases of it, but does not empirically estimate how often it occurs.
\citet{astrauskas20:_how_rust} report that
unsafe Rust code (in their sample set) is typically well-encapsulated.
\citet{qin20:_under_rust} define the notion of ``interior unsafe'' (i.e., a safe function that encapsulates an unsafe function) and find that 58\% of sampled interior unsafe functions in the Rust standard library require conditions to ensure safety but do not check these conditions, relying on being called correctly. Note that this is not a bug---in Rust, the onus is on the caller to ensure documented safety conditions for calling \texttt{unsafe fn}s, and the callee is entitled to ensure that they hold.

It is possible that safety encapsulation holds much more often for arbitrary libraries than for the Rust standard library. That is, the standard library may well
expose more low-level implementation details than other libraries. This question is an important one in terms of understanding how humans use Rust's safety mechanisms, but it is one that needs to be empirically investigated.

\paragraph{Specification language design for unsafety.}
Libraries that expose (rather than encapsulate) safety are required to label relevant functions as
\texttt{unsafe fn}s, and to document (in natural language) the required safety conditions. \texttt{unsafe fn}s must be called from
\texttt{unsafe~\{\}} blocks in the caller. The meaning of
\texttt{unsafe fn} is that the function contract includes
(safety-related) requirements that the caller is responsible for ensuring,
and furthermore, that neither the compiler nor the callee are responsible
for. (In some cases, either the compiler or the callee may fail fast
and refuse to compile or execute non-conforming code.) Dually,
implementations of functions in traits declared as \texttt{unsafe
  trait} must satisfy requirements that the compiler is not responsible for
enforcing. If an Unsafe Rust-using library imposes usage constraints on
its client, and the client violates them, then Rust calls the client's
use of the library \emph{unsound}, and the client has triggered
undefined behaviour. In turn, the client may encapsulate the
unsafety, or it may pass the buck to its own callers, by exposing its
own \texttt{unsafe fn}s.


Verifying that client usages of \texttt{unsafe fn}s are sound requires designers to state, and verifiers to check, some properties that go beyond the expressive capabilities of the Rust type system.
We are not aware of efforts to ensure that C code meets the
required postconditions in analogous cases.


An interesting point for specification language design is that specifications for \texttt{unsafe fn}s may,
somewhat unusually, restrict
what can happen after the function returns, in addition to documenting
(as usual) required preconditions and ensured postconditions. To some extent, these restrictions occur
in other languages too. One example is that in C, after returning from \texttt{free(x)},
the caller must no longer access \texttt{x}. Similarly, in Rust,
value \texttt{x} that has been through \texttt{ManuallyDropped<t>::drop(x)} must no longer be accessed.
As another example, somewhat analogous to the C \texttt{restrict} qualifier, the \texttt{as\_mut} implementation on \texttt{pointer} returns a
value that Rust will assume is unique, but that may actually have aliases;
its postcondition
requires that there be no accesses to the returned value through other
pointers (which the type system does \emph{not} prevent from existing) while the returned value remains in scope.



In the absence of safety encapsulation and in the presence of Unsafe Rust-using libraries, even
verifying the absence of undefined behaviour may require arbitrary
reasoning power and properties; the reason for using \texttt{unsafe} is that the library developer
believes that the required reasoning power is more than what Rust's type system provides.

The proposed verification efforts for Rust will require some thought
about human aspects of verification, namely a notation that allows
library developers to formally specify the needed postconditions, and that
client code can be verified against.
\cite{cui2024unsafe} identified 19 classes of safety properties that various parts of standard library API documentation specified as requirements for callers to unsafe APIs.
We believe that this notation
could well be specialized for the use case of safety properties, and
might turn out to be unnecessary for verifying general properties of
Rust code.

\paragraph{Beyond generic safety properties.}
So far, we have discussed generic safety properties
guaranteed by a lack of undefined behaviour, such as memory safety and
data race freedom. These properties ought to hold in all programs, and
violations of these properties mean that a program is not guaranteed
to satisfy any properties at all. However, a lack of undefined
behaviour is not enough.

Another way for Rust programs to go wrong is by panicking.  For
instance, Rust programs panic when they call {\tt unwrap()} on a {\tt
  None} value.  Prusti~\cite{astrauskas2022prusti} verifies that Rust
programs do not panic due to integer overflows---overflows do not
panic on release builds and are explicitly listed in the Rust
documentation as not being undefined behaviour, but are likely to be
undesired behaviour.  Prusti apparently does not aim to verify
absence of all panics, and for instance, does not detect unwrapping a
{\tt None}.

Classically, program verification efforts consider module or function
preconditions and postconditions (Hoare triples), as well as
invariants; we have mentioned safety-related postconditions above,
and there are also safety-related preconditions.
More generally, there can be functional correctness properties, which
we refer to as \emph{domain-specific properties}. An example
precondition for \texttt{MyVec} is that calling \texttt{insert()}
with an insertion index is only valid if the insertion index is less
than the current length of the \texttt{MyVec}, promising a panic if
the precondition is violated.
Such properties encode specifications, which may be
implicit or written in natural language. Verifying the Rust standard
library could also be construed to require encoding and verifying
these domain-specific properties, but writing a complete set of properties
is at best an art, and we do not believe that there is consensus
on what this set of properties should encompass.

For many programs,
total correctness is desirable, requiring proof of termination. Researchers have also
developed a vast array of techniques to verify termination of
programs~\cite{zhu21:_termin_analy_tears,cook06:_termin_proof_system_code,cousot12:_abstr_inter_framew_termin}.

\section{Extant Verification Tools and Approaches}
\label{sec:tools}

We survey parts of the software verification landscape that are relevant to our efforts to verify Rust. We focus on Java and C/C++, imperative languages (like Rust) that have mature verification ecosystems, sharing important language features with Rust, while also having certain key differences we can learn from. We also briefly discuss Dafny to touch on the interesting case of languages that have built-in verification features. Then, we discuss some of the many Rust verification efforts by others, including the techniques that existing tools use.

\subsection{Other Languages}
\paragraph{Java}
As defined in the Java Language Specification and Java Virtual Machine Specification, Java guarantees memory safety, though not race freedom (which is highly challenging to ensure statically in the presence of unrestricted aliasing). Featherweight Java ~\cite{igarashi2001featherweight} is an influential work in formally reasoning about Java and enables proofs of Java's type safety.

The real situation for Java is not as straightforward and is closer to Rust than one might think. \citet{evans20:_unsaf_class} discusses Unsafe Java features---which exist and are in the Oracle API documentation but not in the official specifications---and how they are used in practice, ubiquitously, by libraries to achieve better performance. As with Rust, we believe that most developers do stick to ``safe Java'' and that the unsafe uses are encapsulated within libraries. \citet{mastrangelo15:_use_you_own_risk} found that 3\% of top-ranked Java artifacts (as of the date of their study) directly used the version of Unsafe available then, and 25\% of artifacts directly or indirectly depend on unsafe code. One difference with Rust appears to be that Rust libraries impose usage conditions on clients for their uses to be sound, and hence to ensure safety, while it appears to us that Java libraries aim to be unconditionally safe, or in other words, to encapsulate safety.

Nevertheless, almost all Java verifiers focus on the verification of ``safe'' Java code (examples include JBMC~\cite{cordeiro2018jbmc}, KeY~\cite{ahrendt2014key}, and OpenJML~\cite{cok2011openjml}).
To our knowledge, the only tool that acknowledges the existence of unsafe Java code is SafeCheck~\cite{huang19:_safec}, which only slightly mitigates the dangers of unsafe Java by adding run-time checks for safety violations and hence improving diagnosability of crashes caused by unsafe code. It does not carry out any static checking of either generic safety properties or domain-specific properties. This is similar to ASan for C/C++~\cite{serebryany12:_addres} or ERASan for Rust~\cite{min24:_erasan}; these dynamic approaches can terminate programs that attempt to trigger undefined behaviour, but not proactively prevent it before runtime.


We also draw an analogy between Java native code and Rust intrinsics, which pose the same kinds of challenges for program analysis and verification. The approach in Java is generally to hardcode the behaviour of the intrinsics and rely on a consensus agreement on this behaviour. Sometimes it is possible to provide a mock implementation of the intrinsic in Java or Rust, but there is no guarantee that the real implementation is faithful; we cannot see any static way of providing guarantees.


\paragraph{C/C++} Rather than being safe by default, the C language is entirely unsafe, meaning that verifiers, regardless of whether they focus on generic safety or domain-specific properties, must always consider a broader range of undefined behaviour. One of the more popular tools is CBMC~\cite{clarke2004tool}, a bounded model checker, whose strengths align particularly well with verifying unsafe and concurrent C code---it is less reliant on invariants than symbolic execution approaches.

Other tools take the deductive approach to verification of C code, such as VST~\cite{appel2011verified} and VCC~\cite{cohen2009vcc}. VST (Verified Software Toolchain) is a formal verification tool that allows users to produce machine-checkable proofs for programs, focusing on both memory safety and functional correctness. On the other hand, VCC takes a modular approach and translates annotated C code into BoogiePL, which is used to generate verification conditions that can be solved by Z3~\cite{de2008z3}.  VCC also verifies both memory safety and functional correctness, but it places a particular focus on concurrent code. Both tools, like effectively any other deductive tool, have the potentially costly requirement of needing function contracts and loop invariants, but in return they support unbounded verification, allowing for stronger guarantees of correctness.


C++ has additional features (e.g., bounds-checking for certain types, more expressive type system including uniqueness). But these features are easy to override---escape hatches are ubiquitous. It therefore seems that most C++ verification tools are extensions of C tools which support C++ features, but which do not trust the stronger type constraints.


\paragraph{Dafny} Some languages are designed specifically for verification, such as Dafny~\cite{leino23:_progr_proof}. Dafny supports a rich specification language and static verification of specifications. While Dafny can compile to existing languages, we consider Dafny and languages like it to be out of scope in our discussion of verification of code in existing language ecosystems. 

Having discussed Java and C/C++, the situation for Rust is closest to Java's, except that we suspect unsafe code to be much more widespread in Rust (given that its ``safe'' mode is more restrictive than Java's). \citet{mastrangelo15:_use_you_own_risk} found that 25\% of surveyed Java artifacts depended on unsafe code, whereas \citet{evans2020rust} found that over half of surveyed Rust crates depended on unsafe code. Rust's unsafe mode is quite similar to C in its power (e.g., it allows manipulating raw pointers), although unsafe Rust code is easy to identify, unlike C code with undefined behaviour, which may be anywhere. Understanding how other languages deal with these challenges can give us insight into how we can do the same for Rust.

\subsection{Rust Verification Tools and Techniques}
There exist a variety of tools for the verification of Rust code, as well as verification frameworks for the implementation and semantics of the Rust language itself, and we survey them here. Many Rust verification tools aim to check properties of safe Rust code, e.g. lack of panics, assertions, and function contracts/invariants. Other tools verify unsafe Rust code and its usage.

\paragraph{Intermediate representations} A verification tool's workflow normally consists of taking source code annotated with function contracts according to some specification, partly compiling the code (mostly via \verb|rustc|), generating some verification conditions, and solving them (e.g. via an SMT or SAT solver). When partly compiling, the tool is in fact lowering the code to an intermediate representation (IR), typically a mid-level IR (MIR) or, at a lower level, an IR at the level of the LLVM IR.

The LLVM IR is the last IR before an executable is produced. The main advantage of using a lower-level IR like this is that other languages also use it (e.g., C), allowing for the straightforward reuse of verification backends from other languages for Rust. Indeed, this is the case for several tools, such as SMACK~\cite{baranowski2018verifying} and SeaHorn~\cite{gurfinkel2015seahorn}, which were originally developed for verification of C (or more generally LLVM-based) code, but can be used on Rust code due to this shared IR. \citet{baranowski2018verifying} show how to verify a lack of overflows in Rust, while an example of Seahorn usage\footnote{\url{https://project-oak.github.io/rust-verification-tools/using-seahorn/}} shows the verification of assertions embedded in the Rust code. The downside of the LLVM IR approach is, due to being lower-level, that information about the original structure and semantics of the program is not readily available~\cite{vanhattum2022verifying}.

At a higher level, the Kani project by \citet{vanhattum2022verifying} specifically chose to analyze Rust's MIR, which also allowed it to incorporate information encoded in a program's dynamic trait objects. Working at a higher level than the LLVM IR retains more Rust-specific semantic information. MIR is used only by Rust, meaning that reusing backends from other languages becomes a more involved process, but it appears to us that this access to richer semantic information is worthwhile. After all, bridging a Rust-specific IR and a generic backend can be engineered, whereas how to recover lost semantic information from a lower-level IR is unclear, and may be impossible in some cases.

\paragraph{Survey of related tools} Prusti~\cite{astrauskas2022prusti} is a deductive verifier that works at MIR level, though it primarily focuses on the verification of functional correctness of safe code. In particular, it verifies that certain panics do not occur in safe Rust code, as well as user-specified assertions. Creusot~\cite{denis2022creusot} is a more recent deductive verifier, also focusing on safe code. It proposes a specification language, Pearlite. The main difference between these two tools is that Prusti specifies mutable borrows using pledges, as opposed to the prophecies (via the \verb|final| operator) used by Creusot. This means that Creusot can support a wider range of borrowing patterns than Prusti (which does not support, for example, reborrowing in a loop~\cite{denis2022creusot}).

Aeneas~\cite{ho2022aeneas} puts forth a fairly novel approach in that it translates Rust programs into a pure-lambda calculus, which allows for much flexibility in terms of the theorem provers that can be used to reason about these programs. Verus~\cite{lattuada2023verus}, on the other hand, has programmers write proofs and specifications using Rust itself, leveraging the language's own type and borrow checking. This way, it can verify even unsafe code, though it has some trouble with more complex borrowing patterns that Aeneas can handle, such as reborrowing. Moreover, it loses out on the flexibility that Aeneas has in terms of being able to reason about programs with almost any theorem prover.

Kani~\cite{vanhattum2022verifying} uses a bounded model checker, translating code from the MIR into Goto-C, which is then passed to CBMC. Besides incorporating higher-level information from Rust, one of the main features of Kani is the verification of unsafe code, which it achieves in part due to its CBMC backend.

Finally, some tools allow for complete formal verification of subsets of the Rust language itself. For instance, RustBelt~\cite{jung2017rustbelt} developed an extensible soundness proof for $\lambda_{\text{Rust}}$, a formal description of the Rust language, minus certain features deemed superfluous to the proof (e.g. traits). Oxide~\cite{weiss2019oxide} also models the key aspects of Rust, supporting certain new features like non-lexical lifetimes. Incidentally, Rust does not have a formal memory model, so we would not be able to verify against an official memory model even if we wanted to.

Gillian-Rust~\cite{ayoun24:_rust} proposes a promising approach to verification of both safe and unsafe Rust. It uses a separation logic combining features of RustBelt and RustHornBelt~\cite{matsushita22:_rusth} to reason about unsafe Rust, and integrates that logic with Creusot to reason about safe Rust (at much lower cost). RefinedRust~\cite{gaher2024refinedrust} is similar to Gillian-Rust in that it uses a separation logic to produce proofs of correctness for both safe and unsafe code, but differs in that these proofs are foundational (i.e., they are machine-checkable in a general-purpose proof assistant). We believe that Gillian-Rust and RefinedRust can verify soundness of the usage of unsafe libraries from safe code, which appears to be out of scope for RustBelt. We also speculate that these tools will be more powerful than the Kani-based approach that we propose here, but also harder to use (e.g., due to their use of separation logic).



\paragraph{Approaches: bounded model checking versus symbolic execution}
In practice, Kani's bounded model checking approach differs from the deductive approach taken by Prusti and Creusot particularly in its handling of loops. The bounded model checking approach explores all behaviours of the program up to a given bound. This approach proved to be particularly fruitful for exploring behaviours of concurrent C programs. From the verification user's perspective, an advantage of the bounded model checking approach is that it does not require loop invariants to verify loops: the verifier simply explores all executions up to the bound. This approach also does not necessarily sacrifice soundness, at least in cases where a fixed bound can be shown to be sufficient. Of course, cases where this cannot be shown represent a limitation of bounded model checking (e.g., unbounded structures like linked lists which need verification beyond the bounds being checked). Bounded model checking approaches should also be easily modifiable to accept loop invariants, though most tools do not seem to currently support them (but Kani now does). Users of deductive verification approaches can specify loop invariants (which can be difficult) or use automatic loop invariant inference techniques (which do not always work). For these reasons, the user should have the option to choose whether or not to use loop invariants (as is the case with Kani for example)---either approach has non-negligible disadvantages and advantages.



\section{Which Code to Verify}
\label{sec:which-code}

We discuss which part of the Rust ecosystem we should focus our verification efforts on. We can consider the Rust ecosystem to be composed of three main groups of code:

\begin{enumerate}
	\item Client code, i.e., applications written in Rust;
	\item Third-party crates, analogous to libraries in other languages; and
	\item The Rust standard library.
\end{enumerate}

These groups are listed from least to most fundamental to the ecosystem. That is, the standard library can be used by many crates, and these crates can be used by many applications; but not vice-versa. Note that verifying, in isolation, a third-party crate that depends on some part of the standard library without having first verified the standard library decreases our confidence in whatever we have proven about the crate. (While the standard library has the advantage that it is widely used, and thus much less likely to have easily-encountered functional correctness issues, experience has shown that security issues do still crop up in the most widely used codebases.) Thus, even though verification of any part of the ecosystem is important, if one is to prioritize certain parts, it is natural to start with the standard library.

Yet, much of the existing literature focuses on the verification of client code and third-party crates. This was the case, for instance, with Prusti~\cite{astrauskas2022prusti}, Kani~\cite{vanhattum2022verifying} (initially), SMACK~\cite{baranowski2018verifying}, and Creusot~\cite{denis2022creusot}. The few works that attempted to verify the standard library did so on subsets of the Rust language (e.g., RustBelt~\cite{jung2017rustbelt}). Indeed, to our knowledge, the only tool that currently explicitly supports verification of the standard library is Kani, with the recent addition of the \verb|verify-std| feature, though the actual verification effort is just beginning. This further supports a prioritization of the verification of the standard library.

Why have people chosen to verify client code? We speculate that when one's tool handles only Safe Rust, one is likely to get further verifying client code; crates and the standard library may contain more Unsafe Rust, and that is harder to verify. Also, as we have stated earlier, verifying Safe Rust frees one to focus on assertions and contracts (if one assumes the underlying Unsafe Rust is properly encapsulated), rather than worrying about safety properties; this focus may be attractive to some researchers.



Finally, we note that a verification effort has two main outputs: a direct one in the form of a proof for a particular program, and occasionally an indirect one in the form of a new technique. We have already discussed how verifying the standard library maximizes the direct output (yielding a proof that is useful to more programs), but if we wish to maximize the total output, we should also consider the indirect output. Naturally, the areas of the ecosystem that are more challenging to verify would require more innovation to do so. We believe that the challenges that the verification of the standard library presents (at least in terms of memory safety and functional correctness) are practically a superset of those of regular crates and client code (e.g., writing function contracts for intrinsics, which we elaborate on in Section~\ref{sec:challenges}). (This is not completely true, since verifying client code may require techniques seldom-used in library code, and vice versa, but it seems like a defensible starting assumption.) Hence, we believe that focusing verification efforts on the standard library has the potential to yield more fruitful verification techniques, and these new techniques could then be applied to other parts of the ecosystem.




\section{Choosing a Tool and Verification Target}
\label{sec:challenges}

Having discussed some Rust verification tools as well as motivating the verification of the Rust standard library, what remains is to determine which tool would be best as a starting point for this task, as well as specific verification targets within the library.

Our first strict requirement is that the tool must support the verification of unsafe Rust code, as it is highly prevalent in the standard library. We therefore do not consider tools that focus only on safe code, like Creusot~\cite{denis2022creusot}, Prusti~\cite{astrauskas2022prusti} and Aeneas~\cite{ho2022aeneas}.  We are looking for a tool that can handle the full complexity of the standard library (e.g., traits) rather than a subset language; techniques like RustBelt~\cite{jung2017rustbelt} and Oxide~\cite{weiss2019oxide} would need to be extended. Finally, our ideal tool is user-friendly: the standard library is a nontrivially-large project, with around 229K lines of code. The process of writing the specifications and performing the proof would therefore preferably be as lightweight as possible (e.g., with less annotation required and as much automation as possible); in terms of human aspects, trading expressive power for ease-of-use as much as possible.

We believe bounded model checking to be a good choice for our goal. Writing annotations and function contracts can be a labour-intensive process, especially for larger projects. Even if bounded model checking still requires contracts, it can sometimes avoid the challenges of writing invariants, which are needed for approaches using symbolic execution, like Prusti~\cite{astrauskas2022prusti} and Creusot~\cite{denis2022creusot}). It also sidesteps the difficulty of using separation logic, required for techniques like Gillian-Rust~\cite{ayoun24:_rust}, Verus~\cite{lattuada2023verus} and RefinedRust~\cite{gaher2024refinedrust}.

To our knowledge, the only tool that satisfies all of our criteria is Kani~\cite{vanhattum2022verifying}. It also has the added advantage of supporting automated verification of some memory safety properties, which very few tools do.
However, it is possible that the boundedness fundamental to bounded model checking will prevent some proofs from going through, e.g. for verifying linked list implementations: we are trading off power for ease of use and seeing how far the bounded model checking approach can take us. In any case, we will focus on using and extending Kani in our efforts to verify the standard library. The choice of verification tool, and its power-versus-annotation burden tradeoff, is a key human factor to consider for verification efforts. Our work will help future researchers choose appropriate tools for their work, based on the experiences that we will share.

For verifying the Rust standard library, we consider two main classes of challenges:
\begin{enumerate}
	\item Supporting verification of features exclusive to the standard library; and,
	\item Ensuring a complete verification of safety properties of unsafe code (heavily used in the standard library).
\end{enumerate}

For (1), consider intrinsics (e.g., \verb|transmute|, which reinterprets the bits of one type as another type). These functions have specifications given in natural language in the Rust documentation, but their implementations are directly woven into the code by the Rust compiler.

From a verification point of view, intrinsics appear as \verb|extern| functions, whose verification Kani does not currently support. Under assume/guarantee reasoning, to verify these functions, we need to (1) have contracts for them (assume) and (2) establish that the functions indeed implement their contracts (guarantee). Natural-language contracts for intrinsics exist in the Rust documentation; we will translate them into a machine-readable form, seek community consensus on the correctness of our translations, and make them available to the verification tool. To our knowledge, techniques for verifying the implementation of intrinsics do not exist for any language, and that part of the problem is likely out of scope of our effort; we would be satisfied with formal specifications for intrinsics, agreed-on by the Rust community.

For (2), we need to improve the verification of unsafe code, which is highly prevalent in the standard library.
While Kani is one of the few tools that does support the verification of unsafe code, there is still a wide range of undefined behaviour it does not detect (e.g., data races, mutation of immutable data, breaking lifetime-related aliasing rules). As stated in Rust's documentation\footnote{\emph{Rustonomicon}, section 16.2, "Behaviour considered undefined": \url{https://doc.rust-lang.org/reference/behavior-considered-undefined.html}}, any program containing undefined behaviour has no semantics, meaning that guarantees given by Kani may not necessarily hold. Thus, it is clear that extending the range of undefined behaviour that Kani detects is crucial to gain better confidence in Kani's proofs.

As mentioned in Section~\ref{sec:introduction}, Amazon has initiated an effort to verify the Rust standard library, and invites community participation in this effort. They recently released a set of challenges to be solved by the Rust verification community\footnote{\emph{Verify Rust Std Lib}, section 3, ``Challenges'': \url{https://model-checking.github.io/verify-rust-std/challenges.html}}, consisting of specific high-priority parts of the standard library that they would like to see verified, including \verb|transmute| and its uses, inductive data types, raw pointer arithmetic operations, and use of raw pointers in intrinsics. We have chosen to start with the \verb|transmute| challenge, as it seems both reasonably achievable
and is a high-value target. Indeed, \verb|transmute| is heavily used throughout the standard library, while also being highly prone to causing undefined behaviour, such as creating instances of types with invalid states. As per the Rustonomicon\footnote{\emph{Rustonomicon}, section 4.4, ``Transmutes'': \url{https://doc.rust-lang.org/nomicon/transmutes.html}}, ``This is really, truly, the most horribly unsafe thing you can do in Rust''. Specifications for the \verb|transmute| function itself would likely include things like checking that the output of \verb|transmute| is valid with respect to the type the input value has been reinterpreted as.

\section{Human-related Outcomes}
\label{sec:human}

By this point, we have motivated the verification of unsafe code in Rust's standard library, as well as the use of Kani for this task. Here, we discuss outcomes desired from this verification effort, beyond the inherent value of the verification itself. In particular, we focus on the interaction between human behaviour and Rust verification. Part of our proposal includes documenting our verification efforts, reflecting on the resulting learnings about Rust verification, and sharing them as an experience report. We plan on answering questions related to specification languages, developer use of safe and unsafe Rust, as well as Kani and its use.

\paragraph{Specification languages} We hope to learn more about how specification languages come into play, as they are a key part of the verification process that humans interact with. For instance, we will investigate the relationship between safety properties in plain text and formalized specification languages, particularly with respect to the completeness of the plain text properties we encounter (i.e., how often are there missing safety properties that need to be added?). Moreover, in the case of the standard library, what kind of specification language is needed to express functional correctness of safe code? In other words, from a human aspects point of view, what improvements to the specification language can be made to bring the verification of safe code closer to full-functional correctness? On the other hand, what does a specification language for safety properties look like, and how specific must it be to safety properties? 

\paragraph{Use of safe/unsafe} By examining lots of Rust code, we hope to understand the conditions under which developers mark safe code as unsafe, and to what extent. This might happen, for instance, when the developer makes the boundaries of the unsafe block much larger than they need to be, encompassing some safe code in the process. This naturally has ramifications in terms of verification, as verification of safe code can benefit from more assumptions that can be made about the code. After all, that safe code will have passed the rigorous compiler checks for safe code.

On the subject of safety properties, we also seek to understand how often clients fail to respect the explicit safety properties that they are bound to uphold. In other words, how often does API misuse occur, and are there patterns in the nature of the misuse? If some stated safety properties are violated a lot more often by clients than others, then understanding the underlying reason could help in establishing some solution. Perhaps stronger checks should be implemented in the APIs for these properties, or maybe the property is generally poorly communicated in documentation. We fundamentally believe that optimizing the affordances of the specification language---an extremely human aspect of enabling reasoning about the code---will contribute to developers writing code that respects necessary safety properties.

As we are primarily focusing on verification of the standard library, we also seek to understand if there is a meaningful difference in the way unsafe code is used in the standard library as opposed to regular libraries. This is important not only because it sheds some light on a potential threat to generalizability for this project, but also because characterizing these differences (if they exist) can help future verification efforts on the standard library to adjust their approaches accordingly.

\paragraph{Kani} As for Kani specifically, by extensively using and eventually extending Kani, we will gain an understanding of which properties can be verified by a model-checking approach such as Kani, and which ones cannot. We plan to augment its existing documentation with our learnings about its abilities and shortcomings, as well as adding capabilities to Kani as needed.

Finally, it is well-understood that usability (or lack thereof) is a major obstacle for the widespread use of software verification in general. If the tools that verify code are easier to use, then more code will be verified. Following our work, we will have a better understanding of which properties are infeasible to verify via Kani. For each of these difficult properties, we also more generally aim to evaluate the cost of verifying them (i.e., the usability of other approaches that allow their verification), as well as the benefit (i.e., how much the lack of verification of this property results in bugs). After we publicize this information, researchers and practitioners will be able to make more informed decisions about which approaches to use, and which properties are worth verifying.

\section{Conclusion}

In this research proposal, we explored the current Rust verification landscape. We first identified two main classes of properties to verify---generic safety and domain-specific properties---and discussed specific properties from these two classes that are worth verifying. Next, we surveyed verification approaches in C/C++ and Java, and then categorized and compared existing Rust techniques. We discussed which parts of the Rust ecosystem are highest-priority for verification, motivating a focus on the standard library. Finally, we motivated the use of Kani for this verification, exploring the challenges inherent to the verification of the standard library and suggesting ways to solve them.

Our exploration has yielded a set of questions and challenges, including:
\begin{itemize}
	\item verifying sound use of unsafe code in Rust (i.e. do clients respect safety properties required to use the unsafe code);
        \item how far bounded model checking can go in verifying unsafe Rust; and
	\item what specifications should be given to \verb|extern| functions in the standard library.
\end{itemize}
We look forward to discussing these challenges and their solutions with the community.

\begin{acks}                            
  This material is based upon work supported in part by an Amazon Research Award. We thank the HATRA reviewers and Michael Coblenz for detailed reviews and discussions which helped improve this work.
\end{acks}


\bibliography{references}


\begin{thebibliography}{40}


\ifx \showCODEN    \undefined \def \showCODEN     #1{\unskip}     \fi
\ifx \showDOI      \undefined \def \showDOI       #1{#1}\fi
\ifx \showISBNx    \undefined \def \showISBNx     #1{\unskip}     \fi
\ifx \showISBNxiii \undefined \def \showISBNxiii  #1{\unskip}     \fi
\ifx \showISSN     \undefined \def \showISSN      #1{\unskip}     \fi
\ifx \showLCCN     \undefined \def \showLCCN      #1{\unskip}     \fi
\ifx \shownote     \undefined \def \shownote      #1{#1}          \fi
\ifx \showarticletitle \undefined \def \showarticletitle #1{#1}   \fi
\ifx \showURL      \undefined \def \showURL       {\relax}        \fi
\providecommand\bibfield[2]{#2}
\providecommand\bibinfo[2]{#2}
\providecommand\natexlab[1]{#1}
\providecommand\showeprint[2][]{arXiv:#2}

\bibitem[\protect\citeauthoryear{Ahrendt, Beckert, Bruns, Bubel, Gladisch,
  Grebing, H{\"a}hnle, Hentschel, Herda, Klebanov, et~al\mbox{.}}{Ahrendt
  et~al\mbox{.}}{2014}]%
        {ahrendt2014key}
\bibfield{author}{\bibinfo{person}{Wolfgang Ahrendt}, \bibinfo{person}{Bernhard
  Beckert}, \bibinfo{person}{Daniel Bruns}, \bibinfo{person}{Richard Bubel},
  \bibinfo{person}{Christoph Gladisch}, \bibinfo{person}{Sarah Grebing},
  \bibinfo{person}{Reiner H{\"a}hnle}, \bibinfo{person}{Martin Hentschel},
  \bibinfo{person}{Mihai Herda}, \bibinfo{person}{Vladimir Klebanov},
  {et~al\mbox{.}}} \bibinfo{year}{2014}\natexlab{}.
\newblock \showarticletitle{The KeY platform for verification and analysis of
  Java programs}. In \bibinfo{booktitle}{\emph{Verified Software: Theories,
  Tools and Experiments: 6th International Conference, VSTTE 2014, Vienna,
  Austria, July 17-18, 2014, Revised Selected Papers 6}}. Springer,
  \bibinfo{pages}{55--71}.
\newblock


\bibitem[\protect\citeauthoryear{Anderson}{Anderson}{2020}]%
        {anderson20:_rust_compil_model_calam}
\bibfield{author}{\bibinfo{person}{Brian Anderson}.}
  \bibinfo{year}{2020}\natexlab{}.
\newblock \bibinfo{title}{The {Rust} Compilation Model Calamity}.
\newblock
  \bibinfo{howpublished}{\url{https://pingcap.medium.com/the-rust-compilation-model-calamity-1a8ce781cf6c}}.
    (\bibinfo{date}{January} \bibinfo{year}{2020}).
\newblock
\newblock
\shownote{Accessed on 14 July 2024.}


\bibitem[\protect\citeauthoryear{Appel}{Appel}{2011}]%
        {appel2011verified}
\bibfield{author}{\bibinfo{person}{Andrew~W Appel}.}
  \bibinfo{year}{2011}\natexlab{}.
\newblock \showarticletitle{Verified Software Toolchain: (Invited Talk)}. In
  \bibinfo{booktitle}{\emph{European Symposium on Programming}}. Springer,
  \bibinfo{pages}{1--17}.
\newblock


\bibitem[\protect\citeauthoryear{Astrauskas, B{\'\i}l{\`y}, Fiala, Grannan,
  Matheja, M{\"u}ller, Poli, and Summers}{Astrauskas et~al\mbox{.}}{2022}]%
        {astrauskas2022prusti}
\bibfield{author}{\bibinfo{person}{Vytautas Astrauskas}, \bibinfo{person}{Aurel
  B{\'\i}l{\`y}}, \bibinfo{person}{Jon{\'a}{\v{s}} Fiala},
  \bibinfo{person}{Zachary Grannan}, \bibinfo{person}{Christoph Matheja},
  \bibinfo{person}{Peter M{\"u}ller}, \bibinfo{person}{Federico Poli}, {and}
  \bibinfo{person}{Alexander~J Summers}.} \bibinfo{year}{2022}\natexlab{}.
\newblock \showarticletitle{The {Prusti} project: Formal verification for
  {Rust}}. In \bibinfo{booktitle}{\emph{NASA Formal Methods Symposium}}.
  Springer, \bibinfo{pages}{88--108}.
\newblock


\bibitem[\protect\citeauthoryear{Astrauskas, Matheja, Poli, M\"{u}ller, and
  Summers}{Astrauskas et~al\mbox{.}}{2020}]%
        {astrauskas20:_how_rust}
\bibfield{author}{\bibinfo{person}{Vytautas Astrauskas},
  \bibinfo{person}{Christoph Matheja}, \bibinfo{person}{Federico Poli},
  \bibinfo{person}{Peter M\"{u}ller}, {and} \bibinfo{person}{Alexander~J.
  Summers}.} \bibinfo{year}{2020}\natexlab{}.
\newblock \showarticletitle{How do programmers use unsafe {Rust}?}
\newblock \bibinfo{journal}{\emph{Proc. ACM Program. Lang.}}
  \bibinfo{volume}{4}, \bibinfo{number}{OOPSLA}, Article
  \bibinfo{articleno}{136} (\bibinfo{date}{nov} \bibinfo{year}{2020}),
  \bibinfo{numpages}{27}~pages.
\newblock
\urldef\tempurl%
\url{https://doi.org/10.1145/3428204}
\showDOI{\tempurl}


\bibitem[\protect\citeauthoryear{Ayoun, Denis, Maksimović, and Gardner}{Ayoun
  et~al\mbox{.}}{2024}]%
        {ayoun24:_rust}
\bibfield{author}{\bibinfo{person}{{Sacha-Élie} Ayoun},
  \bibinfo{person}{Xavier Denis}, \bibinfo{person}{Petar Maksimović}, {and}
  \bibinfo{person}{Philippa Gardner}.} \bibinfo{year}{2024}\natexlab{}.
\newblock \bibinfo{title}{A hybrid approach to semi-automated Rust
  verification}.
\newblock \bibinfo{howpublished}{\url{https://arxiv.org/pdf/2403.15122}}.
  (\bibinfo{date}{March} \bibinfo{year}{2024}).
\newblock


\bibitem[\protect\citeauthoryear{Baranowski, He, and Rakamari{\'c}}{Baranowski
  et~al\mbox{.}}{2018}]%
        {baranowski2018verifying}
\bibfield{author}{\bibinfo{person}{Marek Baranowski}, \bibinfo{person}{Shaobo
  He}, {and} \bibinfo{person}{Zvonimir Rakamari{\'c}}.}
  \bibinfo{year}{2018}\natexlab{}.
\newblock \showarticletitle{Verifying {Rust} programs with {SMACK}}. In
  \bibinfo{booktitle}{\emph{Automated Technology for Verification and Analysis:
  16th International Symposium, ATVA 2018, Los Angeles, CA, USA, October 7-10,
  2018, Proceedings 16}}. Springer, \bibinfo{pages}{528--535}.
\newblock


\bibitem[\protect\citeauthoryear{Clarke, Kroening, and Lerda}{Clarke
  et~al\mbox{.}}{2004}]%
        {clarke2004tool}
\bibfield{author}{\bibinfo{person}{Edmund Clarke}, \bibinfo{person}{Daniel
  Kroening}, {and} \bibinfo{person}{Flavio Lerda}.}
  \bibinfo{year}{2004}\natexlab{}.
\newblock \showarticletitle{A tool for checking ANSI-C programs}. In
  \bibinfo{booktitle}{\emph{Tools and Algorithms for the Construction and
  Analysis of Systems: 10th International Conference, TACAS 2004, Held as Part
  of the Joint European Conferences on Theory and Practice of Software, ETAPS
  2004, Barcelona, Spain, March 29-April 2, 2004. Proceedings 10}}. Springer,
  \bibinfo{pages}{168--176}.
\newblock


\bibitem[\protect\citeauthoryear{Cohen, Dahlweid, Hillebrand, Leinenbach,
  Moskal, Santen, Schulte, and Tobies}{Cohen et~al\mbox{.}}{2009}]%
        {cohen2009vcc}
\bibfield{author}{\bibinfo{person}{Ernie Cohen}, \bibinfo{person}{Markus
  Dahlweid}, \bibinfo{person}{Mark Hillebrand}, \bibinfo{person}{Dirk
  Leinenbach}, \bibinfo{person}{Micha{\l} Moskal}, \bibinfo{person}{Thomas
  Santen}, \bibinfo{person}{Wolfram Schulte}, {and} \bibinfo{person}{Stephan
  Tobies}.} \bibinfo{year}{2009}\natexlab{}.
\newblock \showarticletitle{VCC: A practical system for verifying concurrent
  C}. In \bibinfo{booktitle}{\emph{Theorem Proving in Higher Order Logics: 22nd
  International Conference, TPHOLs 2009, Munich, Germany, August 17-20, 2009.
  Proceedings 22}}. Springer, \bibinfo{pages}{23--42}.
\newblock


\bibitem[\protect\citeauthoryear{Cok}{Cok}{2011}]%
        {cok2011openjml}
\bibfield{author}{\bibinfo{person}{David~R Cok}.}
  \bibinfo{year}{2011}\natexlab{}.
\newblock \showarticletitle{OpenJML: JML for Java 7 by extending OpenJDK}. In
  \bibinfo{booktitle}{\emph{NASA Formal Methods: Third International Symposium,
  NFM 2011, Pasadena, CA, USA, April 18-20, 2011. Proceedings 3}}. Springer,
  \bibinfo{pages}{472--479}.
\newblock


\bibitem[\protect\citeauthoryear{Cook, Podelski, and Rybalchenko}{Cook
  et~al\mbox{.}}{2006}]%
        {cook06:_termin_proof_system_code}
\bibfield{author}{\bibinfo{person}{Byron Cook}, \bibinfo{person}{Andreas
  Podelski}, {and} \bibinfo{person}{Andrey Rybalchenko}.}
  \bibinfo{year}{2006}\natexlab{}.
\newblock \showarticletitle{Termination Proofs for Systems Code}. In
  \bibinfo{booktitle}{\emph{PLDI}}. \bibinfo{pages}{415--426}.
\newblock


\bibitem[\protect\citeauthoryear{Cordeiro, Kesseli, Kroening, Schrammel, and
  Trtik}{Cordeiro et~al\mbox{.}}{2018}]%
        {cordeiro2018jbmc}
\bibfield{author}{\bibinfo{person}{Lucas Cordeiro}, \bibinfo{person}{Pascal
  Kesseli}, \bibinfo{person}{Daniel Kroening}, \bibinfo{person}{Peter
  Schrammel}, {and} \bibinfo{person}{Marek Trtik}.}
  \bibinfo{year}{2018}\natexlab{}.
\newblock \showarticletitle{JBMC: A bounded model checking tool for verifying
  Java bytecode}. In \bibinfo{booktitle}{\emph{International Conference on
  Computer Aided Verification}}. Springer, \bibinfo{pages}{183--190}.
\newblock


\bibitem[\protect\citeauthoryear{Cousot and Cousot}{Cousot and Cousot}{2012}]%
        {cousot12:_abstr_inter_framew_termin}
\bibfield{author}{\bibinfo{person}{Patrick Cousot} {and}
  \bibinfo{person}{Radhia Cousot}.} \bibinfo{year}{2012}\natexlab{}.
\newblock \showarticletitle{An Abstract Interpretation Framework for
  Termination}. In \bibinfo{booktitle}{\emph{Proceedings of the 39th Annual ACM
  SIGPLAN-SIGACT Symposium on Principles of Programming Languages (POPL
  ’12)}}. \bibinfo{address}{Philadelphia, PA, USA},
  \bibinfo{pages}{245--258}.
\newblock


\bibitem[\protect\citeauthoryear{Cui, Sun, Xu, and Zhou}{Cui
  et~al\mbox{.}}{2024}]%
        {cui2024unsafe}
\bibfield{author}{\bibinfo{person}{Mohan Cui}, \bibinfo{person}{Shuran Sun},
  \bibinfo{person}{Hui Xu}, {and} \bibinfo{person}{Yangfan Zhou}.}
  \bibinfo{year}{2024}\natexlab{}.
\newblock \showarticletitle{Is unsafe an Achilles' Heel? A Comprehensive Study
  of Safety Requirements in Unsafe Rust Programming}. In
  \bibinfo{booktitle}{\emph{Proceedings of the IEEE/ACM 46th International
  Conference on Software Engineering}}. \bibinfo{pages}{1--13}.
\newblock


\bibitem[\protect\citeauthoryear{De~Moura and Bj{\o}rner}{De~Moura and
  Bj{\o}rner}{2008}]%
        {de2008z3}
\bibfield{author}{\bibinfo{person}{Leonardo De~Moura} {and}
  \bibinfo{person}{Nikolaj Bj{\o}rner}.} \bibinfo{year}{2008}\natexlab{}.
\newblock \showarticletitle{Z3: An efficient SMT solver}. In
  \bibinfo{booktitle}{\emph{International conference on Tools and Algorithms
  for the Construction and Analysis of Systems}}. Springer,
  \bibinfo{pages}{337--340}.
\newblock


\bibitem[\protect\citeauthoryear{Denis, Jourdan, and March{\'e}}{Denis
  et~al\mbox{.}}{2022}]%
        {denis2022creusot}
\bibfield{author}{\bibinfo{person}{Xavier Denis},
  \bibinfo{person}{Jacques-Henri Jourdan}, {and} \bibinfo{person}{Claude
  March{\'e}}.} \bibinfo{year}{2022}\natexlab{}.
\newblock \showarticletitle{Creusot: a foundry for the deductive verification
  of {Rust} programs}. In \bibinfo{booktitle}{\emph{International Conference on
  Formal Engineering Methods}}. Springer, \bibinfo{pages}{90--105}.
\newblock


\bibitem[\protect\citeauthoryear{Evans, Campbell, and Soffa}{Evans
  et~al\mbox{.}}{2020}]%
        {evans2020rust}
\bibfield{author}{\bibinfo{person}{Ana~Nora Evans}, \bibinfo{person}{Bradford
  Campbell}, {and} \bibinfo{person}{Mary~Lou Soffa}.}
  \bibinfo{year}{2020}\natexlab{}.
\newblock \showarticletitle{Is Rust used safely by software developers?}. In
  \bibinfo{booktitle}{\emph{Proceedings of the ACM/IEEE 42nd International
  Conference on Software Engineering}}. \bibinfo{pages}{246--257}.
\newblock


\bibitem[\protect\citeauthoryear{Evans}{Evans}{2020}]%
        {evans20:_unsaf_class}
\bibfield{author}{\bibinfo{person}{Ben Evans}.}
  \bibinfo{year}{2020}\natexlab{}.
\newblock \bibinfo{title}{The Unsafe Class: Unsafe at Any Speed}.
\newblock
  \bibinfo{howpublished}{\url{https://blogs.oracle.com/javamagazine/post/the-unsafe-class-unsafe-at-any-speed}}.
    (\bibinfo{date}{May} \bibinfo{year}{2020}).
\newblock


\bibitem[\protect\citeauthoryear{G{\"a}her, Sammler, Jung, Krebbers, and
  Dreyer}{G{\"a}her et~al\mbox{.}}{2024}]%
        {gaher2024refinedrust}
\bibfield{author}{\bibinfo{person}{Lennard G{\"a}her}, \bibinfo{person}{Michael
  Sammler}, \bibinfo{person}{Ralf Jung}, \bibinfo{person}{Robbert Krebbers},
  {and} \bibinfo{person}{Derek Dreyer}.} \bibinfo{year}{2024}\natexlab{}.
\newblock \showarticletitle{Refinedrust: A type system for high-assurance
  verification of Rust programs}.
\newblock \bibinfo{journal}{\emph{Proceedings of the ACM on Programming
  Languages}} \bibinfo{volume}{8}, \bibinfo{number}{PLDI}
  (\bibinfo{year}{2024}), \bibinfo{pages}{1115--1139}.
\newblock


\bibitem[\protect\citeauthoryear{Gurfinkel, Kahsai, Komuravelli, and
  Navas}{Gurfinkel et~al\mbox{.}}{2015}]%
        {gurfinkel2015seahorn}
\bibfield{author}{\bibinfo{person}{Arie Gurfinkel}, \bibinfo{person}{Temesghen
  Kahsai}, \bibinfo{person}{Anvesh Komuravelli}, {and} \bibinfo{person}{Jorge~A
  Navas}.} \bibinfo{year}{2015}\natexlab{}.
\newblock \showarticletitle{The SeaHorn verification framework}. In
  \bibinfo{booktitle}{\emph{International Conference on Computer Aided
  Verification}}. Springer, \bibinfo{pages}{343--361}.
\newblock


\bibitem[\protect\citeauthoryear{Ho and Protzenko}{Ho and Protzenko}{2022}]%
        {ho2022aeneas}
\bibfield{author}{\bibinfo{person}{Son Ho} {and} \bibinfo{person}{Jonathan
  Protzenko}.} \bibinfo{year}{2022}\natexlab{}.
\newblock \showarticletitle{Aeneas: Rust verification by functional
  translation}.
\newblock \bibinfo{journal}{\emph{Proceedings of the ACM on Programming
  Languages}} \bibinfo{volume}{6}, \bibinfo{number}{ICFP}
  (\bibinfo{year}{2022}), \bibinfo{pages}{711--741}.
\newblock


\bibitem[\protect\citeauthoryear{Huang, Guo, Li, Li, Qi, Chow, and Huang}{Huang
  et~al\mbox{.}}{2019}]%
        {huang19:_safec}
\bibfield{author}{\bibinfo{person}{Shiyou Huang}, \bibinfo{person}{Jianmei
  Guo}, \bibinfo{person}{Sanhong Li}, \bibinfo{person}{Xiang Li},
  \bibinfo{person}{Yumin Qi}, \bibinfo{person}{Kingsum Chow}, {and}
  \bibinfo{person}{Jeff Huang}.} \bibinfo{year}{2019}\natexlab{}.
\newblock \showarticletitle{SafeCheck: Safety Enhancement of Java Unsafe API}.
  In \bibinfo{booktitle}{\emph{2019 IEEE/ACM 41st International Conference on
  Software Engineering (ICSE)}}. \bibinfo{pages}{889--899}.
\newblock
\urldef\tempurl%
\url{https://doi.org/10.1109/ICSE.2019.00095}
\showDOI{\tempurl}


\bibitem[\protect\citeauthoryear{Igarashi, Pierce, and Wadler}{Igarashi
  et~al\mbox{.}}{2001}]%
        {igarashi2001featherweight}
\bibfield{author}{\bibinfo{person}{Atsushi Igarashi},
  \bibinfo{person}{Benjamin~C Pierce}, {and} \bibinfo{person}{Philip Wadler}.}
  \bibinfo{year}{2001}\natexlab{}.
\newblock \showarticletitle{Featherweight Java: a minimal core calculus for
  Java and GJ}.
\newblock \bibinfo{journal}{\emph{ACM Transactions on Programming Languages and
  Systems (TOPLAS)}} \bibinfo{volume}{23}, \bibinfo{number}{3}
  (\bibinfo{year}{2001}), \bibinfo{pages}{396--450}.
\newblock


\bibitem[\protect\citeauthoryear{Jung, Jourdan, Krebbers, and Dreyer}{Jung
  et~al\mbox{.}}{2017}]%
        {jung2017rustbelt}
\bibfield{author}{\bibinfo{person}{Ralf Jung}, \bibinfo{person}{Jacques-Henri
  Jourdan}, \bibinfo{person}{Robbert Krebbers}, {and} \bibinfo{person}{Derek
  Dreyer}.} \bibinfo{year}{2017}\natexlab{}.
\newblock \showarticletitle{{RustBelt}: Securing the foundations of the {Rust}
  programming language}.
\newblock \bibinfo{journal}{\emph{Proceedings of the ACM on Programming
  Languages}} \bibinfo{volume}{2}, \bibinfo{number}{POPL}
  (\bibinfo{year}{2017}), \bibinfo{pages}{1--34}.
\newblock


\bibitem[\protect\citeauthoryear{Klabnik and Nichols}{Klabnik and
  Nichols}{2024}]%
        {klabnik24:_rust_progr_languag}
\bibfield{author}{\bibinfo{person}{Steve Klabnik} {and} \bibinfo{person}{Carol
  Nichols}.} \bibinfo{year}{2024}\natexlab{}.
\newblock \bibinfo{booktitle}{\emph{The {Rust} Programming Language}}.
\newblock \bibinfo{publisher}{No Starch Press}.
\newblock
\newblock
\shownote{Accessed at \url{https://doc.rust-lang.org/book/} on 14 July 2024.}


\bibitem[\protect\citeauthoryear{Lattuada, Hance, Cho, Brun, Subasinghe, Zhou,
  Howell, Parno, and Hawblitzel}{Lattuada et~al\mbox{.}}{2023}]%
        {lattuada2023verus}
\bibfield{author}{\bibinfo{person}{Andrea Lattuada}, \bibinfo{person}{Travis
  Hance}, \bibinfo{person}{Chanhee Cho}, \bibinfo{person}{Matthias Brun},
  \bibinfo{person}{Isitha Subasinghe}, \bibinfo{person}{Yi Zhou},
  \bibinfo{person}{Jon Howell}, \bibinfo{person}{Bryan Parno}, {and}
  \bibinfo{person}{Chris Hawblitzel}.} \bibinfo{year}{2023}\natexlab{}.
\newblock \showarticletitle{Verus: Verifying rust programs using linear ghost
  types}.
\newblock \bibinfo{journal}{\emph{Proceedings of the ACM on Programming
  Languages}} \bibinfo{volume}{7}, \bibinfo{number}{OOPSLA1}
  (\bibinfo{year}{2023}), \bibinfo{pages}{286--315}.
\newblock


\bibitem[\protect\citeauthoryear{Leino}{Leino}{2023}]%
        {leino23:_progr_proof}
\bibfield{author}{\bibinfo{person}{K.~Rustan~M. Leino}.}
  \bibinfo{year}{2023}\natexlab{}.
\newblock \bibinfo{booktitle}{\emph{Program Proofs}}.
\newblock \bibinfo{publisher}{MIT Press}.
\newblock


\bibitem[\protect\citeauthoryear{Leroy}{Leroy}{2009}]%
        {Leroy-Compcert-CACM}
\bibfield{author}{\bibinfo{person}{Xavier Leroy}.}
  \bibinfo{year}{2009}\natexlab{}.
\newblock \showarticletitle{Formal verification of a realistic compiler}.
\newblock \bibinfo{journal}{\emph{Commun. ACM}} \bibinfo{volume}{52},
  \bibinfo{number}{7} (\bibinfo{year}{2009}), \bibinfo{pages}{107--115}.
\newblock
\urldef\tempurl%
\url{http://xavierleroy.org/publi/compcert-CACM.pdf}
\showURL{%
\tempurl}


\bibitem[\protect\citeauthoryear{Mastrangelo, Ponzanelli, Mocci, Lanza,
  Hauswirth, and Nystrom}{Mastrangelo et~al\mbox{.}}{2015}]%
        {mastrangelo15:_use_you_own_risk}
\bibfield{author}{\bibinfo{person}{Luis Mastrangelo}, \bibinfo{person}{Luca
  Ponzanelli}, \bibinfo{person}{Andrea Mocci}, \bibinfo{person}{Michele Lanza},
  \bibinfo{person}{Matthias Hauswirth}, {and} \bibinfo{person}{Nathaniel
  Nystrom}.} \bibinfo{year}{2015}\natexlab{}.
\newblock \showarticletitle{Use at You Own Risk: The {Java} Unsafe {API} in the
  Wild}. In \bibinfo{booktitle}{\emph{OOPSLA 2015: Proceedings of the 2015 ACM
  SIGPLAN International Conference on Object-Oriented Programming, Systems,
  Languages, and Applications}}. \bibinfo{pages}{695--710}.
\newblock


\bibitem[\protect\citeauthoryear{Matsushita, Denis, Jourdan, and
  Dreyer}{Matsushita et~al\mbox{.}}{2022}]%
        {matsushita22:_rusth}
\bibfield{author}{\bibinfo{person}{Yusuke Matsushita}, \bibinfo{person}{Xavier
  Denis}, \bibinfo{person}{Jacques-Henri Jourdan}, {and} \bibinfo{person}{Derek
  Dreyer}.} \bibinfo{year}{2022}\natexlab{}.
\newblock \showarticletitle{RustHornBelt: a semantic foundation for functional
  verification of Rust programs with unsafe code}. In
  \bibinfo{booktitle}{\emph{Proceedings of the 43rd ACM SIGPLAN International
  Conference on Programming Language Design and Implementation}}
  \emph{(\bibinfo{series}{PLDI 2022})}. \bibinfo{publisher}{Association for
  Computing Machinery}, \bibinfo{address}{New York, NY, USA},
  \bibinfo{pages}{841–856}.
\newblock
\showISBNx{9781450392655}
\urldef\tempurl%
\url{https://doi.org/10.1145/3519939.3523704}
\showDOI{\tempurl}


\bibitem[\protect\citeauthoryear{Min, Yu, Jeong, Song, and Jeon}{Min
  et~al\mbox{.}}{2024}]%
        {min24:_erasan}
\bibfield{author}{\bibinfo{person}{Jiun Min}, \bibinfo{person}{Dongyeon Yu},
  \bibinfo{person}{Seongyun Jeong}, \bibinfo{person}{Dokyung Song}, {and}
  \bibinfo{person}{Yuseok Jeon}.} \bibinfo{year}{2024}\natexlab{}.
\newblock \showarticletitle{ERASAN : Efficient Rust Address Sanitizer}. In
  \bibinfo{booktitle}{\emph{2024 IEEE Symposium on Security and Privacy (SP)}}.
  \bibinfo{publisher}{IEEE Computer Society}, \bibinfo{address}{Los Alamitos,
  CA, USA}, \bibinfo{pages}{239--239}.
\newblock
\showISSN{2375-1207}
\urldef\tempurl%
\url{https://doi.org/10.1109/SP54263.2024.00258}
\showDOI{\tempurl}


\bibitem[\protect\citeauthoryear{Ozdemir}{Ozdemir}{2019}]%
        {ozdemir19:_unsaf_rust}
\bibfield{author}{\bibinfo{person}{Alex Ozdemir}.}
  \bibinfo{year}{2019}\natexlab{}.
\newblock \bibinfo{title}{Unsafe in {Rust}: The {Abstraction Safety Contract}
  and Public Escape}.
\newblock
  \bibinfo{howpublished}{\url{https://cs.stanford.edu/~aozdemir/blog/unsafe-rust-escape}}.
    (\bibinfo{year}{2019}).
\newblock


\bibitem[\protect\citeauthoryear{Qin, Chen, Yu, Song, and Zhang}{Qin
  et~al\mbox{.}}{2020}]%
        {qin20:_under_rust}
\bibfield{author}{\bibinfo{person}{Boqin Qin}, \bibinfo{person}{Yilun Chen},
  \bibinfo{person}{Zeming Yu}, \bibinfo{person}{Linhai Song}, {and}
  \bibinfo{person}{Yiying Zhang}.} \bibinfo{year}{2020}\natexlab{}.
\newblock \showarticletitle{Understanding memory and thread safety practices
  and issues in real-world {Rust} programs}. In
  \bibinfo{booktitle}{\emph{Proceedings of the 41st ACM SIGPLAN Conference on
  Programming Language Design and Implementation}} \emph{(\bibinfo{series}{PLDI
  2020})}. \bibinfo{publisher}{Association for Computing Machinery},
  \bibinfo{address}{New York, NY, USA}, \bibinfo{pages}{763–779}.
\newblock
\showISBNx{9781450376136}
\urldef\tempurl%
\url{https://doi.org/10.1145/3385412.3386036}
\showDOI{\tempurl}


\bibitem[\protect\citeauthoryear{{Rust team}}{{Rust team}}{2016}]%
        {rustonomicon}
\bibfield{author}{\bibinfo{person}{{Rust team}}.}
  \bibinfo{year}{2016}\natexlab{}.
\newblock \bibinfo{title}{The {Rustonomicon}: What Unsafe Means}.
\newblock
  \bibinfo{howpublished}{\url{https://doc.rust-lang.org/nomicon/safe-unsafe-meaning.html}}.
    (\bibinfo{year}{2016}).
\newblock
\newblock
\shownote{Accessed: 2024-07-12.}


\bibitem[\protect\citeauthoryear{Serebryany, Bruening, Potapenko, and
  Vyukov}{Serebryany et~al\mbox{.}}{2012}]%
        {serebryany12:_addres}
\bibfield{author}{\bibinfo{person}{Konstantin Serebryany},
  \bibinfo{person}{Derek Bruening}, \bibinfo{person}{Alexander Potapenko},
  {and} \bibinfo{person}{Dmitriy Vyukov}.} \bibinfo{year}{2012}\natexlab{}.
\newblock \showarticletitle{AddressSanitizer: A fast address sanity checker}.
  In \bibinfo{booktitle}{\emph{2012 USENIX annual technical conference (USENIX
  ATC 12)}}. \bibinfo{pages}{309--318}.
\newblock


\bibitem[\protect\citeauthoryear{Shirzad and Lam}{Shirzad and Lam}{2024}]%
        {shirzad24:_rust}
\bibfield{author}{\bibinfo{person}{Mohammad~Robati Shirzad} {and}
  \bibinfo{person}{Patrick Lam}.} \bibinfo{year}{2024}\natexlab{}.
\newblock \showarticletitle{A study of common bug fix patterns in {Rust}}.
\newblock \bibinfo{journal}{\emph{Empirical Software Engineering}}
  \bibinfo{volume}{29}, \bibinfo{number}{2} (\bibinfo{date}{February}
  \bibinfo{year}{2024}).
\newblock


\bibitem[\protect\citeauthoryear{VanHattum, Schwartz-Narbonne, Chong, and
  Sampson}{VanHattum et~al\mbox{.}}{2022}]%
        {vanhattum2022verifying}
\bibfield{author}{\bibinfo{person}{Alexa VanHattum}, \bibinfo{person}{Daniel
  Schwartz-Narbonne}, \bibinfo{person}{Nathan Chong}, {and}
  \bibinfo{person}{Adrian Sampson}.} \bibinfo{year}{2022}\natexlab{}.
\newblock \showarticletitle{Verifying dynamic trait objects in {Rust}}. In
  \bibinfo{booktitle}{\emph{Proceedings of the 44th International Conference on
  Software Engineering: Software Engineering in Practice}}.
  \bibinfo{pages}{321--330}.
\newblock


\bibitem[\protect\citeauthoryear{Weiss, Gierczak, Patterson, and Ahmed}{Weiss
  et~al\mbox{.}}{2019}]%
        {weiss2019oxide}
\bibfield{author}{\bibinfo{person}{Aaron Weiss}, \bibinfo{person}{Olek
  Gierczak}, \bibinfo{person}{Daniel Patterson}, {and} \bibinfo{person}{Amal
  Ahmed}.} \bibinfo{year}{2019}\natexlab{}.
\newblock \showarticletitle{Oxide: The essence of {Rust}}.
\newblock \bibinfo{journal}{\emph{arXiv preprint arXiv:1903.00982}}
  (\bibinfo{year}{2019}).
\newblock


\bibitem[\protect\citeauthoryear{Xu, Chen, Sun, Zhou, and Lyu}{Xu
  et~al\mbox{.}}{2021}]%
        {xu21:_memor}
\bibfield{author}{\bibinfo{person}{Hui Xu}, \bibinfo{person}{Zhuangbin Chen},
  \bibinfo{person}{Mingshen Sun}, \bibinfo{person}{Yangfan Zhou}, {and}
  \bibinfo{person}{Michael~R Lyu}.} \bibinfo{year}{2021}\natexlab{}.
\newblock \showarticletitle{Memory-safety challenge considered solved? An
  in-depth study with all {Rust} {CVEs}}.
\newblock \bibinfo{journal}{\emph{ACM Transactions on Software Engineering and
  Methodology (TOSEM)}} \bibinfo{volume}{31}, \bibinfo{number}{1}
  (\bibinfo{date}{September} \bibinfo{year}{2021}), \bibinfo{pages}{1--25}.
\newblock
\urldef\tempurl%
\url{https://doi.org/10.1145/3466642}
\showDOI{\tempurl}


\bibitem[\protect\citeauthoryear{Zhu and Kincaid}{Zhu and Kincaid}{2021}]%
        {zhu21:_termin_analy_tears}
\bibfield{author}{\bibinfo{person}{Shaowei Zhu} {and} \bibinfo{person}{Zachary
  Kincaid}.} \bibinfo{year}{2021}\natexlab{}.
\newblock \showarticletitle{Termination Analysis without the Tears}. In
  \bibinfo{booktitle}{\emph{PLDI}}.
\newblock


\end{thebibliography}



\end{document}